# GENERALISATION OF THE TWO-SCALE MOMENTUM THEORY FOR COUPLED WIND TURBINE/FARM OPTIMISATION

Takafumi NISHINO[1)]

ABSTRACT

An extended theoretical approach is proposed to predict the average power of wind turbines in a large finite-size wind farm. The approach is based on the two-scale momentum theory proposed recently for the modelling of ideal very large wind farms, but the theory is now generalised by introducing the effect of additional pressure difference induced by the farm between the upstream and downstream sides of the farm, making the approach applicable to real wind farms that are large but not as large as the size of the relevant atmospheric system driving the flow over the farm. To validate the generalised theoretical model, 3D Reynolds-averaged Navier-Stokes simulations of boundary layer flow over a large array (25 × 25) of actuator (drag) discs are conducted at eight different conditions. The results suggest that the generalised model could be embedded in a regional-scale atmospheric model to predict the average power of a given wind farm effectively. This approach may also be combined with the blade element momentum (BEM) theory for coupled wind turbine/farm optimisation.

## 1. INTRODUCTION

Wind energy is one of the most common forms of natural energy that are readily available on the Earth. The total installed capacity of both onshore and offshore wind farms has dramatically increased worldwide over the last decade, and the size of each individual wind farm has also been increasing significantly in recent years. For example, in 2016 DONG Energy (now Ørsted) announced the outline of 1.2GW Hornsea Project One wind farm, which is currently under construction and expected to be fully operational by 2021 with up to 174 large (7MW) wind turbines covering a large offshore area of approximately 400 square kilometres off the coast of Yorkshire in northern England.

A series of recent theoretical and computational studies by the author and his collaborators[1-4] has shown that the power generation characteristics of wind turbines in a very large wind farm may be systematically analysed using a coupled turbine-scale and farm-scale momentum conservation argument. More recently, we[5] proposed a combination of the original 'two-scale coupled' momentum theory for a very large array of ideal wind turbines[1] with the classical blade element momentum (BEM) theory for turbine rotor design, making it possible to explore a potentially important relationship between the design of turbine rotors and their performance in a very large wind farm. A major assumption employed in these recent studies using the two-scale coupled momentum conservation argument, however, is that the

1)    Departmental Lecturer in Civil Engineering Fluid Mechanics, Department of Engineering Science, University of Oxford, Parks Road, Oxford OX1 3PJ, United Kingdom momentum loss (or pressure drop) from the most upstream position to the most downstream position of the wind farm site is unchanged before and after farm construction; this essentially means that the farm is assumed to be as large as the relevant atmospheric system that drives the flow over the farm[5]. In order for the two-scale momentum theory (and its combination with the BEM theory for coupled turbine/farm optimisation) to be applicable to real wind farms of different sizes, this major assumption about the farm size needs to be relaxed.

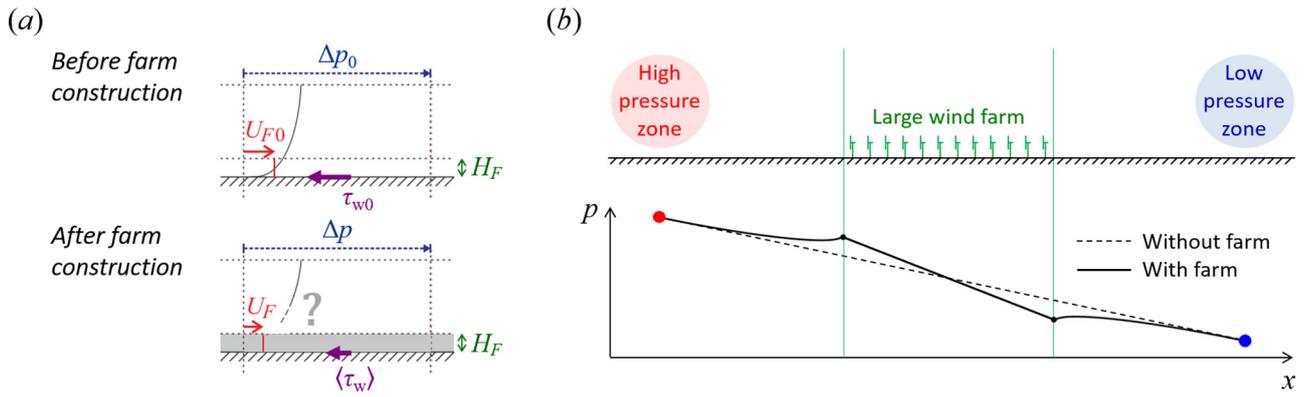

Figure 1: Schematic of the generalised two-scale momentum theory: (*a*) fully developed boundary layers observed inside the farm site before and after farm construction; (*b*) additional pressure difference induced by the farm.

In this short paper, I present a simple but effective extension of the two-scale momentum theory for the modelling and optimisation of large wind farms. This extension makes it possible to predict the power generation characteristics of wind turbines not only in the special 'very large' farm situation considered in the original theory but also in more general situations where the farm is large but not as large as the relevant atmospheric system; therefore, the extended theory may be referred to as the 'generalised' two-scale momentum theory. A series of CFD simulations of boundary layer flow over a large array (25 × 25) of actuator discs (representing simplified wind turbines) is also presented for comparison with the theory, followed by discussion and conclusions.

## 2. THEORY

The original two-scale momentum theory was proposed by Nishino[1] in 2016 to derive the lowest order (quasi-1D) flow model that predicts the performance of very large wind farms in a highly simplified manner. Although the final form of the derived model (hereafter referred to as 'N2016' model) is very simple, the underlying theory (or the set of concepts and assumptions) is not so simple and readers are referred to Section 2.1 of the recent paper by Nishino and Hunter[5] for full details. As with the original theory, here we consider a number of identical and periodically arranged wind turbines forming a large wind farm, the size of which is large enough to assume that the flow through the farm is mostly fully developed (this 'large enough' size may depend significantly on the atmospheric stability[6] but is likely to be of the order of 10km). The only difference from the original theory, however, is that now we do not assume a fixed streamwise pressure gradient driving the flow over the farm but take into account that this pressure gradient may change before and after farm construction, as depicted in Fig. 1 (note that the pressure here is still considered to vary linearly from the most upstream position to the most downstream position of the farm as the flow through the farm is assumed to be fully developed). Hence, by considering the momentum balance equation for the fully developed flow before and after farm construction separately and then coupling the two equations, we obtain

$$\langle \tau_w \rangle S + T = \frac{\Delta p}{\Delta p_0} \tau_{w0} S \ , \tag{1}$$

where $\tau_{w0}$ and $\langle \tau_w \rangle$ are the (horizontally-averaged) wind-induced shear stress on the land or sea surface observed before and after farm construction, respectively; $T$ is the thrust on each turbine in the farm; $S$ is the land/sea surface area per turbine; and $\Delta p_0$ and $\Delta p$ are the pressure drop (across each area $S$, for example) observed before and after farm construction, respectively (note that $\Delta p = \Delta p_0$ in the original theory). Following the original theory, the above equation can be transformed into the following 'two-scale coupled' momentum balance equation:

$$4\alpha(1-\alpha)\frac{\Lambda}{C_{f0}}\beta^2 + \beta^\gamma - 1 = \frac{(\Delta p - \Delta p_0)}{\Delta p_0} \ , \tag{2}$$

where $\alpha$ and $\beta$ are the turbine-scale and farm-scale flow reduction factors defined as $\alpha = U_T/U_F$ and $\beta = U_F/U_{F0}$ (with $U_T$ and $U_F$ representing the wind speed averaged over the turbine's swept area $A$ and over the volume $H_F S$, respectively, where $H_F$ is the nominal farm-layer height defined in the original theory and is typically two to three times as high as the turbine hub height, and $U_{F0}$ being the farm-layer-averaged wind speed observed before farm construction); $\Lambda$ is the farm density defined as $\Lambda = A/S$; $C_{f0}$ is the natural friction coefficient of the land/sea surface defined as $C_{f0} = \tau_{w0}/\frac{1}{2}\rho U_{F0}^2$ (with $\rho$ being the air density); and $\gamma$ is the empirical model parameter defined as $\gamma = \log_\beta(\langle\tau_w\rangle/\tau_{w0})$, the value of which is typically around 1.5 and less than 2. It should be noted that the only difference of the above equation from the original N2016 model is in its right-hand side, namely the farm-induced pressure term, which now takes a non-zero value (unless $\Delta p = \Delta p_0$). If the value of this additional pressure term is known, then this equation can be solved to obtain $\beta$ as a function of $\alpha$ (for a given set of parameters: $\Lambda$, $C_{f0}$ and $\gamma$) and thereby the power coefficient of each turbine in the farm as $C_P = T U_T/\frac{1}{2}\rho U_{F0}^3 A = 4\alpha^2(1-\alpha)\beta^3$. By considering a range of $\alpha$, we can find the optimal flow reduction factors ($\alpha_{\text{opt}}$ and $\beta_{\text{opt}}$) to maximise the value of $C_P$.

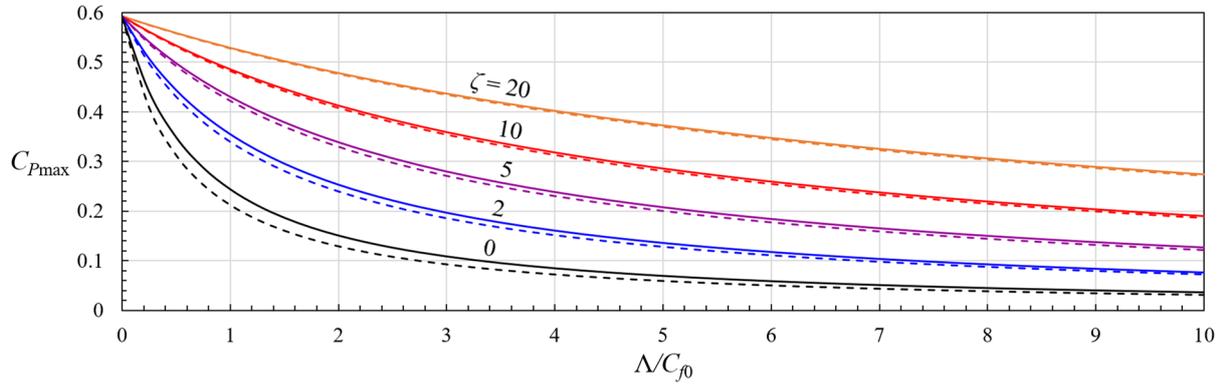

Figure 2: Maximum power coefficient ($C_{P\max}$) predicted by the generalised two-scale momentum theory (solid lines: $\gamma = 2$; dashed lines: $\gamma = 1.5$); note that $\zeta = 0$ corresponds to the original two-scale momentum theory.

The value of the farm-induced pressure term in Eq. (2) may depend on many environmental (i.e. atmospheric and geographical) conditions, including the ratio of the farm size to the size of the relevant atmospheric system driving the flow (the latter could be of the order of 10km to 1,000km, depending on the type of wind generation mechanism); but for a given environment, this pressure term should depend primarily on the farm-scale flow reduction factor $\beta$. In fact, the results of CFD simulations presented in the next section will show that the farm-induced pressure tends to increase approximately linearly with the farm-scale flow induction factor $(1-\beta)$; hence, we may model this pressure term as

$$\frac{(\Delta p - \Delta p_0)}{\Delta p_0} = \zeta(1-\beta) \ , \tag{3}$$

where $\zeta$ is an 'environment-dependent' parameter and needs to be determined empirically for a given wind farm. As will be discussed later, this model parameter may be determined using a numerical atmospheric model for the site of interest with a single volume of momentum sink representing a whole wind farm. Figure 2 shows the maximum power coefficient ($C_{P\max}$) predicted by the above 'generalised' two-scale momentum model, plotted against the normalised farm density ($\Lambda/C_{f0}$). As can be seen, the generalised model predicts that the performance of a large wind farm may depend significantly on $\zeta$ and $\Lambda/C_{f0}$, whereas the effect of $\gamma$ is relatively minor (especially when $\zeta$ is large).

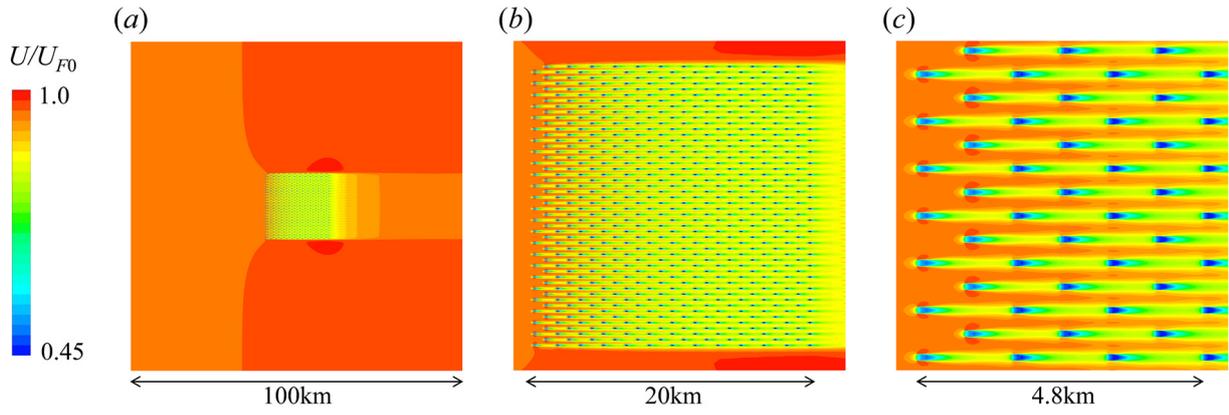

Figure 3: Streamwise velocity contours at the hub height ($z/d = 1$) for the higher bottom roughness case with $C'_T = 1.5$: (a) entire computational domain; (b, c) enlarged views of the farm and the first seven rows.

## 3. CFD ANALYSIS

A series of 3D incompressible Reynolds-averaged Navier-Stokes (RANS) simulations of boundary layer flow over a staggered array of 625 actuator discs (momentum sinks) has been conducted using a commercial finite-volume CFD solver ANSYS FLUENT 16.2. All 625 discs are arranged periodically with a fixed streamwise ($x$) and spanwise ($y$) spacing of $8d$, where $d = 100$m is the disc diameter, representing a large wind farm consisting of 25 rows of 25 turbines over a 20-kilometre square area as shown in Fig. 3. The farm is located horizontally at the centre of a large rectangular computational domain of $(L_x \times L_y \times L_z) = (1,000d \times 1,000d \times 10d)$, and the vertical distance from the bottom boundary (representing the land/sea surface) to the centre of each disc (corresponding to the turbine hub) is $1d$. Each disc is modelled as a stationary permeable surface, on which the momentum loss (per unit area) is calculated as $M_x = \frac{1}{2}\rho U^2 K$, where $U$ is the local (not disc-averaged) streamwise velocity and $K$ is the resistance coefficient or the momentum loss factor (given as an input); therefore, the power coefficient of each disc is $C_P = \int M_x U \, dA / \frac{1}{2}\rho U_{F0}^3 A = K \int U^3 \, dA / U_{F0}^3 A$. It should be noted that $K$ is in general slightly different from the local thrust coefficient defined as $C'_T = T / \frac{1}{2}\rho U_T^2 A$ (since $U_T^2 A$ is slightly different from $\int U^2 \, dA$ unless $U$ is perfectly uniform over the surface of each disc as in the quasi-1D theory); however, this difference tends to be very small (typically less than 1%) and hence, we may consider that $K$ and $C'_T$ are equivalent. The RANS equations are solved with the standard $k$-$\varepsilon$ model of Launder and Spalding but with a local suppression of turbulent viscosity around the edge of each disc as in the previous study[1] (this is to ensure that the axial induction factor of a single 'isolated' disc would agree well with the classical actuator disc theory, i.e. $a = (1 - \alpha) = C'_T/(C'_T + 4)$, for $0 \le C'_T \le 2$; see the original paper[1] for further details). The density and viscosity of the working fluid (air) are set to 1.225 kg/m³ and 1.789×10⁻⁵ kg/m-s, respectively.

The bottom boundary of the computational domain is treated in the same manner as in the study by Zapata et al.[2] using a modified wall function approach for 'fully rough' surfaces available in ANSYS FLUENT 16.2. Two different roughness levels (corresponding to two aerodynamic roughness lengths: $z_0 = 50$mm and 0.5mm) are considered in the present study; these two cases (high $z_0$ and low $z_0$) can be taken as examples of onshore and offshore environments, respectively. The top and side boundaries of the domain are treated as symmetry boundaries, whereas the downstream boundary is defined as a pressure outlet (i.e. pressure is fixed whilst all other flow variables are extrapolated from the interior of the domain). The upstream boundary is defined as a velocity inlet, where fully developed profiles of $U$, $k$ and $\varepsilon$ (obtained from a precursor, fully developed boundary layer simulation without actuator discs for each $z_0$ case) are prescribed. Of importance here is the mass flow rate (or the average velocity) given at the inlet. Since the purpose of this CFD analysis is to validate the generalised two-scale momentum theory, ideally, the mass flow rate should be adjusted in such a way that the pressure drop from the inlet to the outlet of this 100km domain (obtained as a result of each farm simulation) will be the same as that in a corresponding 'without farm' simulation (required to obtain $U_{F0}$ and $\Delta p_0$ for comparison with the theory); this is to mimic possible 'before and after construction' situations where the size of a given atmospheric system driving the flow is 100km. However, such an adjustment of the mass flow rate in the farm simulation tends to make it difficult to obtain fully converged solutions. Hence, all farm simulations in this study are conducted with a fixed mass flow rate (with an inlet-average velocity of 10m/s) and instead, the mass flow rate in the corresponding 'without farm' case is adjusted after each farm simulation (thereby obtaining different values of $U_{F0}$ and $\Delta p_0$ for each farm simulation). This approach is justified since the change in the mass flow rate is small enough (only up to 7% in this study) to assume that any Reynolds-number effects due to that change are insignificant.

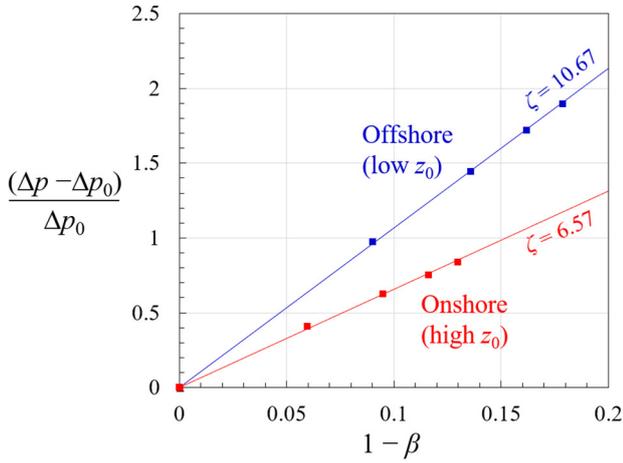

Figure 4: Relationship between the normalised pressure difference induced by the farm and the farm-scale flow induction factor (symbols: CFD results for $C_T' = 0.5$, 1, 1.5 and 2; lines: linear fits of the CFD results).

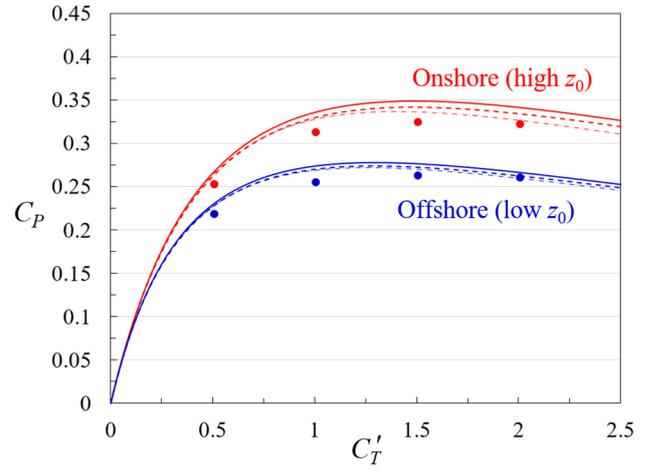

Figure 5: Comparison of the power coefficient between the CFD (symbols) and theory (lines); solid and dashed: $\gamma = 2$ and 1.5, respectively, with a linear fit for $\frac{(\Delta p - \Delta p_0)}{\Delta p_0}$; dash-dot: $\gamma = 1.5$ with a quadratic fit for $\frac{(\Delta p - \Delta p_0)}{\Delta p_0}$.

The computational grid employed (for both 'farm' and 'without farm' simulations) is a multi-block structured grid composed of 19.2 million hexahedral cells. The 2D cross-sectional ($y$-$z$) mesh topology around each disc is similar to that in the previous study[1] but the resolution is coarser (with only 24 grid points along the edge of each disc compared to 64 grid points in the previous study). The minimum cell thickness along the disc edge is $0.02d$, whereas that along the bottom boundary is $0.1d$. For the third direction, the minimum and maximum cell sizes are $0.04d$ (immediately upstream and downstream of each disc) and $5d$ (away from the farm), respectively. The discretisation scheme used is nominally second-order accurate, and less than 10,000 iterations are required to obtain fully converged results.

Figure 4 shows the relationship between the farm-induced pressure difference and the farm-scale flow induction factor obtained from four 'onshore' and four 'offshore' farm simulations (with $C_T' = 0.5$, 1.0, 1.5 and 2.0 for each case). The nominal farm-layer height $H_F$ was found to be $2.6d$ in all cases; hence $U_F$ and $U_{F0}$ (to obtain $\beta = U_F/U_{F0}$) were calculated as volume averages over a rectangular farm-layer zone of ($200d \times 200d \times 2.6d$). Meanwhile, $\Delta p$ and $\Delta p_0$ were calculated as pressure drops (across the entire farm length) averaged over the entire farm width ($200d$) and the entire domain height ($10d$), e.g. $\Delta p = \frac{1}{200d \times 10d} \int_0^{10d} \int_{-100d}^{100d} (p_{\text{up}} - p_{\text{down}}) \, \mathrm{d}y \, \mathrm{d}z$, where $p_{\text{up}}(y, z)$ and $p_{\text{down}}(y, z)$ are the local pressures at the most upstream position ($x = -100d$, which is $4d$ upstream of the first row of discs) and most downstream position ($x = 100d$, which is $4d$ downstream of the 25th row of discs) of the farm site, respectively. As can be seen, the relationship is approximately linear and the gradient $\zeta$ can be determined by using a linear fit of the results for each $z_0$ case. Figure 5 compares predictions of the (average) power coefficient between the CFD and the generalised two-scale momentum theory. The CFD results plotted here are averages of all 625 discs, whereas the theoretical results are obtained with key model-input parameters determined from the CFD results ($\Lambda/C_{f0} = 2.29$ and $\zeta = 6.57$ for onshore; $\Lambda/C_{f0} = 5.56$ and $\zeta = 10.67$ for offshore). Note that $C_T' = 4(1 - \alpha)/\alpha$ in the theory[5]. It can be seen that, for both onshore and offshore cases, the CFD results agree well with the theoretical model. The agreement improves further, but only slightly, if we employ a quadratic (instead of linear) fit to model the farm-induced pressure term. Figure 6 shows variations of the row-averaged $C_P$ for the onshore and offshore cases with $C_T' = 1.5$. As expected from the velocity contours shown earlier in Fig. 3, the first two rows (where discs are not in the wake of other discs) yield much higher $C_P$ values than the downstream rows, but again the theory well predicts the farm-averaged $C_P$.

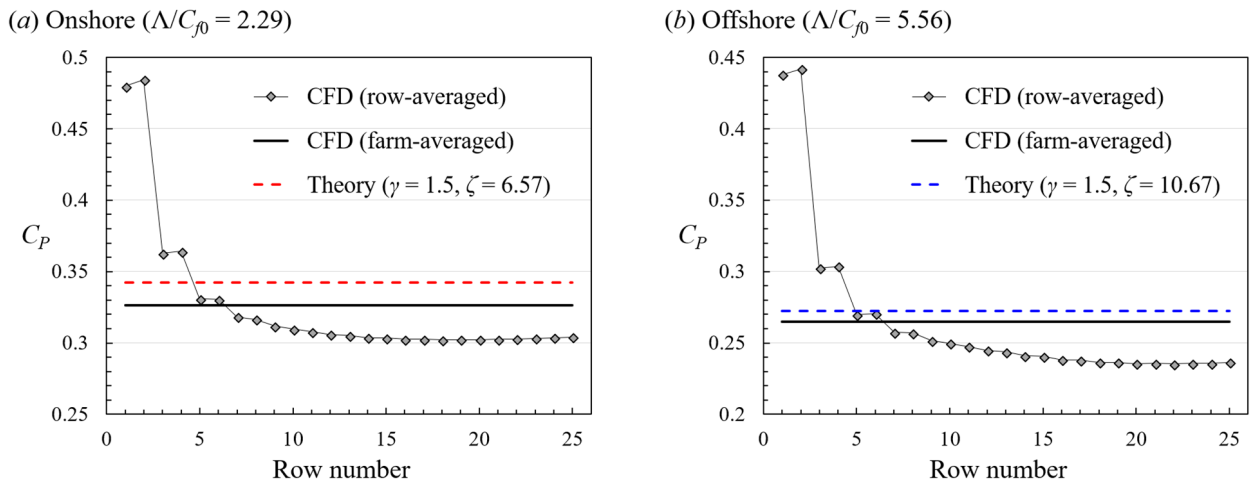

Figure 6: Variation of the row-averaged power coefficient from the first row to the last (25th) row of the farm with $C_T' = 1.5$: (*a*) higher bottom roughness case; (*b*) lower bottom roughness case.

## 4. DISCUSSION AND CONCLUSIONS

The results presented above indicate that the (average) power characteristics of wind turbines in a large wind farm may be predicted instantly as a function of the local thrust coefficient $C_T'$ if the environment-dependent parameter $\zeta$ in the proposed theoretical model is known. To examine this point further, additional lower-resolution CFD simulations of the same pressure-driven flow over a single volume of momentum sink representing the entire farm (instead of 625 individual discs) have been conducted (not shown here due to the limited space) and these simplified simulations have yielded $\zeta$ values close to those obtained from the simulations presented earlier. This suggests that, in future studies, the proposed model could be used in conjunction with a regional-scale numerical atmospheric model (in which the whole farm may be modelled as a single momentum sink) to quickly predict the performance of a given large wind farm.

In conclusion, the two-scale momentum theory proposed earlier[1] has been generalised by introducing the effect of farm-induced pressure modelled as a function of the farm-scale flow induction factor. The generalised theory can still be easily combined with the BEM theory (as in the previous study[5]) for coupled wind turbine/farm optimisation.